\documentclass[12pt]{iopart}

\usepackage{subfigure}
\usepackage{graphicx}
\usepackage{color}
\begin{document}

\title[Thermodynamic uncertainty relation in a tilted periodic potential under coarse graining]{Thermodynamic uncertainty relation in a tilted periodic potential under coarse graining}
\author{Takaaki Monnai}
\address{Department of Materials and Life Science, Seikei University, Tokyo, 180-8633, Japan}
\ead{monnai@st.seikei.ac.jp}
\vspace{10pt}
\begin{indented}
\item[]July 2021
\end{indented}

\begin{abstract}
Recently, some general relations have been studied in nonequilibrium mesoscopic systems. 
In particular, the thermodynamic uncertainty relation (TUR) provides a universal  internal relation among the cumulants of currents and the entropy production. 
In this paper, we give a simple derivation of TUR for thermally fluctuating  particle current in a tilted periodic potential from a coarse grained point of view with the use of the transition probabilities. Also, we explore the condition for the bound of TUR to monotonically increase with the affinity. 
\end{abstract}

%
%
%
%
%

\section{Introduction}
Recent advances in nano-technology such as the fabrication and manipulation of nano devices\cite{Fujisawa1,Fujisawa2,Pekola1} demands new theories for the nonequilibrium mesoscopic systems.
The mesoscopic systems are large compared with the atomic scale but sufficiently small so that fluctuation substantially affect the dynamics. 
For example, the fluctuation theorem focuses on the probability distribution of the entropy production, which satisfies a model independent symmetry of the large deviation function\cite{Gallavotti1,Kurchan1,Lebowitz1,Jarzynski1,Crooks1,Crooks2,Harbola1,Andrieux1,Rao1}. 

Here, our interest lies in another general theory of the stochastic thermodynamics, i.e., the so-called thermodynamic uncertainty relation (TUR) $\frac{\langle J^2\rangle_c}{\langle J\rangle^2}\geq\frac{2}{\langle\sigma\rangle}$  among the mean $\langle J\rangle$ and the variance $\langle J^2\rangle_c$ of the current $J$ and the expectation value of the entropy production $\langle\sigma\rangle$\cite{Barato1,Gingurich1}. It implies that the precision $\frac{\langle J \rangle^2}{\langle J^2\rangle_c}$ is bounded by the mean entropy production $\langle\sigma\rangle$, which are independently measurable, and therefore TUR is of practical importance. 

There are several derivations of TUR for examples on the basis of large deviation theory\cite{Gingurich1,Gingurich2,Pietzonka1}, the information geometry\cite{Goold1}, and the Riesz representation theorem\cite{Falasco1}. There are also many generalizations and related works\cite{Hasegawa1,Pietzonka1,Pietzonka2,Horowitz1,Polettini1}. 

It is instructive to illustrate the underlying physical insight by a simple model. 
One of the simplest models describing the nonequilibrium steady state (NESS) is the thermal diffusion in a tilted periodic potential\cite{Reimann1,Reimann2}. 
This model plays a role in many phenomena such as the diffusion of ions on crystal surfaces\cite{Frenken1}, the phase diffusion of the Josephson junction\cite{Ankerhold1}, and transports in biophysical systems\cite{Reimann2}.   
In the context of the stochastic thermodynamics, there is a transparent derivation of TUR for this model\cite{Hyeon1} from microscopic point of view in terms of the overdamped Langevin equation with a microscopic time resolution.
TUR bound was rigorously obtained in particular for the sufficiently small and large tilting regimes. 

In this paper, we give another contribution to TUR for the tilted periodic potential by showing a simple kinetic derivation under a proper coarse graining procedure with a finite time resolution\cite{Monnai1,Monnai2}.  
We also investigate under which condition the bound for TUR is tight in terms of the Kramers escape rate\cite{van Kampen1}. 
  
This paper is organized as follows.
In Sec. 2, we describe our model.
In Sec. 3, we derive TUR in terms of the Kramers transition rate. 
Sec. 4 is devoted to a summary. 
\section{Model}
Let us consider the thermal diffusion of a particle on an $L$-periodic potential $V(x+L)=V(x)$ dragged by a constant load force $F$. We assume that the particle satisfies an overdamped Langevin equation
\begin{eqnarray}
&&\gamma\frac{d}{dt}x(t)=-\frac{d}{dx}V_{eff}(x)+\xi(t), \label{stochastic1}
\end{eqnarray}
where $\gamma$ and $V_{eff}(x)=V(x)-Fx$ denote the friction coefficient and the effective potential. And, the thermal noise $\xi(t)$ satisfies the fluctuation-dissipation relation $\langle\xi(t)\xi(t')\rangle=2\frac{\gamma }{\beta}\delta(t-t')$. Here, $\beta=\frac{1}{k_BT}$ stands for the inverse temperature. 
In order to calculate the Kramers escape rate, we explore the probability density. 
The corresponding Fokker-Planck equation for the probability density is
\begin{eqnarray}
&&\frac{\partial}{\partial t}P(x,t)=-\frac{\partial}{\partial x}J(x,t), \label{distribution1}
\end{eqnarray}
with the probability current 
\begin{eqnarray}
&&J(x,t) \nonumber \\
&=&-\frac{1}{\gamma}\frac{dV_{eff}(x)}{dx}P(x,t)-\frac{1}{\beta\gamma}\frac{\partial}{\partial x}P(x,t). \label{current1}
\end{eqnarray}
We consider (\ref{distribution1}) under the periodic boundary condition $P(x+L,t)=P(x,t)$.
In NESS, the probability current becomes constant\cite{Reimann2}
\begin{eqnarray}
J^{st}=N(1-e^{-\beta FL}) \label{steady1}
\end{eqnarray}
with 
\begin{eqnarray}
N=\frac{1}{\beta\gamma}\left(\int_0^L dx\int_x^{x+L}dye^{\beta(V_{eff}(y)-V_{eff}(x))}\right)^{-1}. \label{coefficient1}
\end{eqnarray}
We assume that there exists one minimum $x_{min}$ and one maximum $x_{max}$ of effective potential in each period $L$, and the noise is weak compared with the barrier height $\frac{1}{\beta}\ll\Delta V_{eff}=V_{eff}(x_{max})-V_{eff}(x_{min})$.
Then, the saddle point approximation gives the following expression of the stationary current
\begin{eqnarray}
&&J^{st}=k_+-k_- \label{current3}
\end{eqnarray}
 in terms of the Kramers escape rates
\begin{eqnarray}
&&k_+=\frac{|\frac{d^2}{dx^2}V_{eff}(x_{min})\frac{d^2}{dx^2}V_{eff}(x_{max})|^{\frac{1}{2}}}{2\pi\gamma}e^{-\beta\Delta V_{eff}} \nonumber \\
&&k_-=k_+e^{-\beta FL}. \label{escape1}
\end{eqnarray}
Here, $k_+$ and $k_-$ stand for the Kramers escape rates, the probability per unit time for a particle escape to the right and left minima, respectively. 
In particular, the local detailed balance condition is satisfied
\begin{eqnarray}
&&\frac{k_+}{k_-}=e^{\beta FL}. \label{local1}
\end{eqnarray}
\section{Derivation of TUR}
In terms of the transition rates, we can take a coarse grained picture for the stochastic motion of the particle as a biased random walk among the potential minima. Let us suppose that the particle moves to the right and the left minima with probabilities $p_+\propto k_+$ and $p_-\propto k_-$ at discrete time $t=0,\tau,2\tau,...$ with a time step $\tau$. 
\begin{figure}
\center{
\includegraphics[scale=0.8]{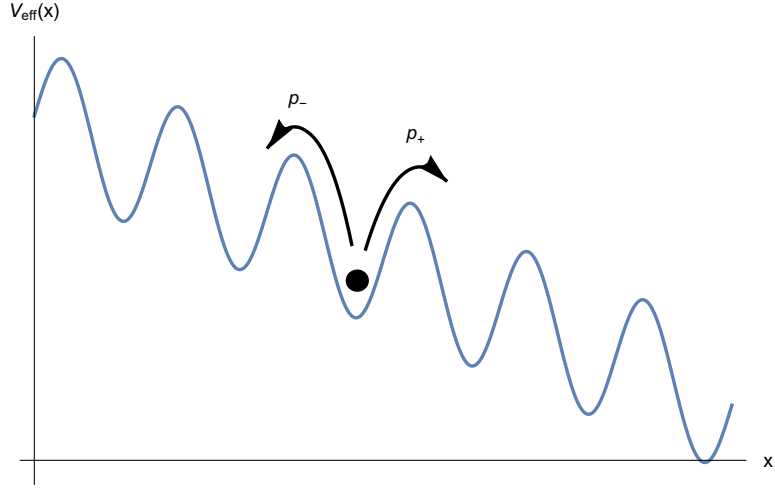}
}
\caption{Schematic illustration of the tilted periodic potential. The particle moves to the right and left with transition probabilities $p_+$ and $p_-$ at $t=m\tau$ $m=0,1,2,...$.}
\end{figure}        
After $n$ transitions, the probability that the particle moves $kL$ to the right, $lL$ to the left, and remains in the same minimum $m=n-(k+l)$ times in total is 
\begin{eqnarray}
&&\pi(k,l)=\frac{n!}{k!l!(n-(k+l))!}p_+^kp_-^l(1-(p_++p_-))^{n-k-l}. \label{probability1}
\end{eqnarray}
First, we calculate the mean of the particle current as 
\begin{eqnarray}
&&\langle J\rangle \nonumber \\
&=&\sum_{k+l+m=n}(k-l)\frac{n!}{k!l!m!}p_+^kp_-^l(1-(p_++p_-))^m \nonumber \\
&=&(p\frac{\partial}{\partial p}-q\frac{\partial}{\partial q})(p+q+r)^n|_{p=p_+,q=p_-,r=1-(p_++p_-)} \nonumber \\
&=&n(p_+-p_-). \label{mean1}
\end{eqnarray}
Here, the sum runs all $k,l,m\geq 0$ that satisfy $k+l+m=n$. 
Similarly, we can calculate the variance of the particle current as 
\begin{eqnarray}
&&\langle J^2\rangle_c \nonumber \\
&=&\sum_{k+l+m=n}(k-l-\langle J\rangle)^2\frac{n!}{k!l!m!}p_+^kp_-^l(1-(p_++p_-))^m \nonumber \\
&=&n(p_+-p_+^2+2p_+p_-+p_--p_-^2), \label{variance1}
\end{eqnarray}
which is proportional to the time duration. This feature of the cumulant is important. 

On the other hand, the mean entropy production is calculated from the Joule heat as 
\begin{eqnarray}
&&\langle\sigma\rangle=n(p_+-p_-)\beta FL. \label{entropy1}
\end{eqnarray}

From (\ref{local1}), the transition probabilities also satisfy the local detailed balance condition
\begin{eqnarray}
&&\frac{p_+}{p_-}=e^{\beta FL}. \label{local2}
\end{eqnarray}
Therefore, the ratio between the variance and the square of the mean of the current is given as
\begin{eqnarray}
&&\frac{\langle J^2\rangle_c}{\langle J\rangle^2} \nonumber \\
&=&\frac{1}{n(p_+-p_-)}\biggl(\frac{1+e^{-\beta FL}}{1-e^{-\beta FL}}-p_+(1-e^{-\beta FL})\biggr). \label{ratio1}    
\end{eqnarray}

Let us explore TUR by multiplying $\langle\sigma\rangle$ to (\ref{ratio1})
\begin{eqnarray}
&&\frac{\langle J^2\rangle_c}{\langle J\rangle^2}\langle\sigma\rangle \nonumber \\
&=&\beta FL\left(\frac{1+e^{-\beta FL}}{1-e^{-\beta FL}}-p_+(1-e^{-\beta FL})\right). \label{TUR1}
\end{eqnarray}
It is straightforward to show that the function in the right hand side of (\ref{TUR1}) 
\begin{eqnarray}
&&f(x,p_+) \nonumber \\
&=&x\left(\frac{1+e^{-x}}{1-e^{-x}}-p_+(1-e^{-x})\right) \nonumber \\
&=&2+(\frac{1}{6}-p_+)x^2+\frac{p_+}{2}x^3+{\cal O}(x^4) \label{function1}
\end{eqnarray}
is positive for $p_+\leq 1$ and larger than $2$ for all $x$ with $p_+$ being sufficiently small. Fix the load force $F$ and the temperature, the smallness of $p_+$ corresponds to the choice of a short time step $\tau$. This implies the TUR for the tilted periodic potential under the coarse graining. 

Fig. 1 shows that the minimum value of (\ref{TUR1}) $\displaystyle\min_{\beta FL}\frac{\langle J^2\rangle_c}{\langle J^2\rangle}\langle\sigma\rangle$ by varying $p_+$. 
It implies that for $p_+\ll 1$ the minimum of (\ref{TUR1}) is almost equal to $2$. 
The presence of this plateau region shows that the TUR bound is tight for short $\tau$ as in \cite{Hyeon1}.   
\begin{figure}
\center{
\includegraphics[scale=0.8]{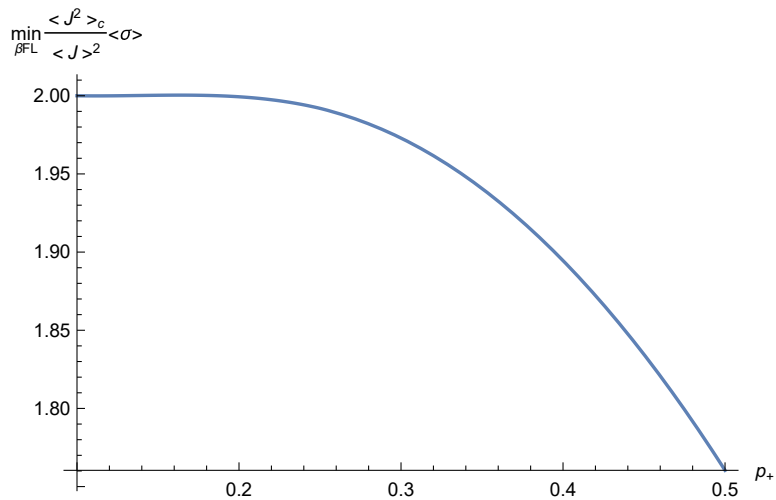}
}
\caption{The minimum of the ratio between the precision and the mean entropy production $\displaystyle\min_{\beta FL}\frac{\langle J^2\rangle_c}{\langle J\rangle^2}\langle\sigma\rangle$ as a function of $p_+$. For $p_+\ll 1$ under a short $\tau$, there is a plateau where the minimum of the ratio is almost equal to $2$.} 
\end{figure}
In Fig. 2, we illustrated $\frac{\langle J^2\rangle_c}{\langle J\rangle^2}\langle\sigma\rangle$ as a function of $\beta FL$ for $p_+\ll 1$ (shaded region), and $p_+=0.5$ (blue) and $p_+=0.7$ (red). If either the load force $F$ is sufficiently strong or the time step $\tau$ is long enough then the precision can be large for the intermediate values of $\beta FL$, since the fluctuation of the particle current is suppressed.     
\begin{figure}
\center{
\includegraphics[scale=0.8]{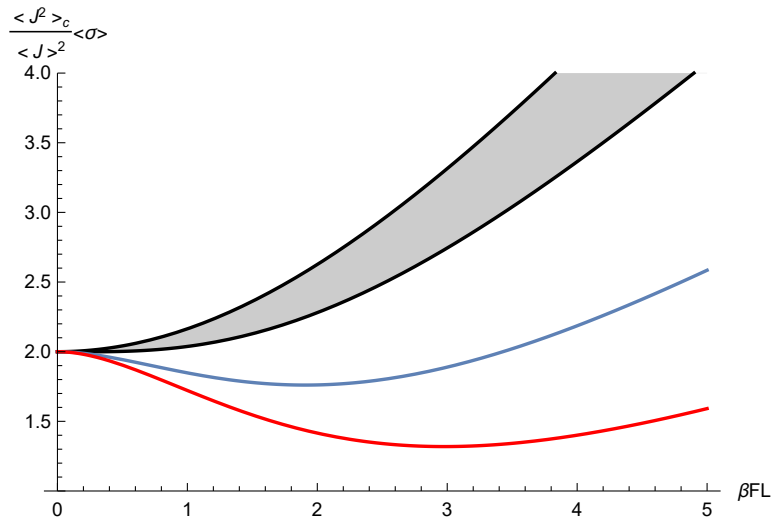}
}
\caption{We illustrate TUR ratio $\frac{\langle J^2\rangle_c}{\langle J\rangle^2}\langle\sigma\rangle$ as a function of $\beta FL$ for $p_+\ll 1$ with a sufficiently short time step $\tau$ (shaded region), and $p_+=0.5$ (blue) and $p_+=0.7$ (red).}
\end{figure}

\section{Summary}
We straightforwardly derived TUR for the particle current of the tilted periodic potential under a coarse graining. The point is that we used the description of a biased random walk   satisfying the local detailed balance condition. In this manner, we can explicitly evaluate the mean and the variance of the current, and also calculate the entropy production from the Joule heat.  
In particular, the variance of the current is proportional to the number of time steps owing to the additivity.      
The local detailed balance condition is inherited from that of the Kramers escape rates by restricting to the weak noise regime. 

We also investigated the ratio between the precision $\frac{\langle J^2\rangle}{\langle J^2\rangle_c}$ and the mean entropy production $\langle\sigma\rangle$, which is almost equal to $2$ as in Fig. 1 for $p_+\ll 1$ and is monotonically increasing function of the dimensionless quantity $\beta FL$ for $p_+\ll 1$ shown as shaded region in Fig. 2. Note that our result is consistent with that of \cite{Hyeon1}, which corresponds to the short enough time resolution $\tau$ and thus $p_+\ll 1$. On the other hand, the ratio has a local minimum for $p_+\gg\frac{1}{6}$ with a sufficiently long time step $\tau$, which connotes that the precision can be large for the intermediate strength of the affinity $\beta FL$. Also, the ratio is larger than $2$ for large enough values of $\beta FL$, which is actually realized for a fixed $F$ by taking $\beta$ large. 
\section{Acknowledgment}
This work was supported by the Grant-in-Aid for Scientific Research (C) (No.~18K03467) from the Japan Society for the Promotion of Science (JSPS).  
\appendix
\section{Cumulants} 
In Sec. 2, we calculated the mean and the variance of the particle current. Here, we explore the higher order cumulants. As shown in \cite{Monnai2}, the characteristic function of $\sigma=\beta FL k$ for the net displacement $L k$ is given by 
\begin{eqnarray}
&&\log\langle e^{-\lambda\sigma}\rangle \nonumber \\
&=&\log(\sum_{l=0}^n\sum_{k+l=0}^{n-l}\frac{n!}{(k+l)!l!(n-l-2l)!}\left(\frac{p_+}{p_-}\right)^{\lambda k}p_+^{k+l}p_-^l(1-(p_++p_-))^{n-k-2l}) \nonumber \\
&=&n\log(1-p_+(1-e^{-\lambda\beta FL})+p_-(e^{\lambda\beta FL}-1)). \label{characteristic}
\end{eqnarray}
It is evident that all the cumulants are proportional to the total number of transitions $n$. 
Then, the $m$-th order cumulant divided by $m!$ in the expansion of (\ref{characteristic}) is negligible for large $m$. This point is compatible to TUR given by the lowest two terms $-\langle\sigma\rangle+\frac{1}{2}\langle\sigma^2\rangle_c$ in the cumulant expansion of (\ref{characteristic}) with $\lambda=1$.  

For example, we can reproduce the mean and the variance, and the third order cumulant is given as
\begin{eqnarray}
&&\langle(\sigma-\langle\sigma\rangle)^3\rangle \nonumber \\
&=&n(p_+-p_--3(p_+^2-p_-^2)+2(p_+^3-p_-^3)-6p_+p_-(p_+-p_-)). \label{skewness1}
\end{eqnarray}
The higher order cumulants can be calculated as well.

\end{document}